\documentclass[final,5p,times,authoryear]{elsarticle}

\usepackage{amsmath,amsfonts}
\usepackage{algorithmic}
\usepackage{array}
\usepackage{subfig}
\usepackage{textcomp}
\usepackage{url}
\usepackage{verbatim}
\usepackage{graphicx}
\usepackage{bbm}
\usepackage{multirow}
\usepackage{booktabs}
\usepackage{rotating}
\usepackage{colortbl}
\usepackage{xcolor}
\usepackage{diagbox}
\usepackage{makecell}
\usepackage{xurl} 

\usepackage{microtype}
\definecolor{mypink}{RGB}{255,105,180}

\usepackage[
  colorlinks=true,
  citecolor=blue,   
  linkcolor=mypink, 
  urlcolor=mypink   
]{hyperref}

\definecolor{prismblue}{RGB}{230,240,255}
\newcommand{\etal}{\textit{et~al.}}

\journal{Speech Communication}
\biboptions{authoryear,round}

\begin{document}

\begin{frontmatter}

\title{DPDFNet: Boosting DeepFilterNet2 via Dual-Path RNN}

\author[inst1]{Daniel Rika\fnref{fn1}\corref{cor1}}
\ead{Daniel.Rika@ceva-ip.com}

\author[inst1]{Nino Sapir\fnref{fn1}}
\ead{Nino.Sapir@ceva-ip.com}

\author[inst1]{Ido Gus}
\ead{Ido.Gus@ceva-ip.com}

\address[inst1]{Ceva Technologies, Ltd, 7 Hapnina Street, Ra'anana 4321545, Israel}

\cortext[cor1]{Corresponding author.}

\fntext[fn1]{D.~Rika and N.~Sapir contributed equally to this work.}

\begin{abstract}
We present DPDFNet, a causal single-channel speech enhancement model that extends DeepFilterNet2 architecture with dual-path blocks in the encoder, strengthening long-range temporal and cross-band modeling while preserving the original enhancement framework. In addition, we demonstrate that adding a loss component to mitigate over-attenuation in the enhanced speech, combined with a fine-tuning phase tailored for ``always-on'' applications, leads to substantial improvements in overall model performance.
We evaluate DPDFNet on the standard VoiceBank+DEMAND and DNS4 blind test benchmarks, where it shows consistent gains over DeepFilterNet2 and strong overall performance against other causal open-source models. In addition, we introduce a supplementary multilingual low-SNR evaluation set comprising long recordings in 12 languages across everyday noise scenarios, on which DPDFNet delivers superior performance to other causal open-source models, including some that are substantially larger and more computationally demanding. We also propose an holistic metric named PRISM, a composite, scale-normalized aggregate of intrusive and non-intrusive metrics, which demonstrates clear scalability with the number of dual-path blocks.
We further demonstrate on-device feasibility by deploying DPDFNet on Ceva-NeuPro\texttrademark-Nano edge NPUs. Results indicate that DPDFNet-4, our second-largest model, achieves real-time performance on NPN32 and runs even faster on NPN64, confirming that state-of-the-art quality can be sustained within strict embedded power and latency constraints.
\end{abstract}

\begin{keyword}
Speech enhancement\sep Noise reduction\sep Dual-Path RNN\sep Real-time systems\sep Embedded systems
\end{keyword}

\end{frontmatter}


\section{Introduction}
\label{sec:intro}

Single-channel speech enhancement (SE) aims to recover clear, natural speech from a noisy recording when only one microphone is available, a common setting in telephony, conferencing, mobile devices, and assistive products. Unlike multi-microphone arrays, the single-channel case has no spatial cues from additional microphones and must rely on time and frequency information, which makes reliable operation at low SNRs especially hard. For interactive use, models must run \emph{causally} with low delay to avoid conversational lag and audio-video drift, and they should remain stable across changes in noise, rooms, and languages. While large, non-causal systems can excel offline, there is a growing need for methods that deliver high perceived quality when run in a streaming setting.

Early real-time speech enhancement often combined traditional signal processing with compact neural networks. Valin \citep{rnnoise} proposed RNNoise, which predicts critical-band gains from features such as Bark-frequency cepstral coefficients and pitch correlations, then applies a frequency-domain pitch filter. This hybrid design achieves very low latency and embedded-class efficiency, and it outperforms minimum mean-square spectral estimators in perceptual quality.
Research then advanced sequence modeling for streaming noise suppression. Westhausen and Meyer \citep{DTLN} introduced DTLN with two cascaded cores. One core operates in the frequency-domain and the other uses a learned time-domain representation. Both rely on LSTM layers with instant layer normalization, enabling stable frame-synchronous processing with fewer than one million parameters and benefiting from both magnitude and phase cues. Braun and Tashev proposed NSNet2 \citep{nsnet2}, a recurrent model that estimates spectral suppression gains from log power spectral features. Its level-invariant loss normalization and broad augmentation over SNR, spectral shape, and signal level yield a robust real-time baseline.
Another line of work moved to the waveform domain. Defossez et al. \citep{DEMUCS} proposed DEMUCS, an encoder-decoder with U-Net-style skip connections and a recurrent bottleneck that operates directly on raw audio. Training uses a mix of waveform loss and multi-resolution spectral losses with strong augmentation. This approach improves naturalness and intelligibility under real-time constraints. Kong et al. proposed CleanUNet \citep{CleanUNet} as a refinement of this design. It strengthens the bottleneck with stacked masked self-attention and combines waveform and spectral objectives, including a high-band loss that improves the handling of silence and high-frequency detail. Although relatively large, CleanUNet maintains low latency and delivers consistent quality.
Work in the time-frequency domain focused on stronger cross-band modeling while preserving local structure. Hao et al. \citep{Fullsubnet} proposed FullSubNet, which encodes long-range context with a full-band LSTM. The frame-wise embedding is concatenated with local frequency neighborhoods and passed to a shared sub-band LSTM that predicts a complex ideal ratio mask \citep{cIRM} (cIRM) with tanh compression. This design links global spectral trends with local stationarity cues.
Dual-path recurrent models were optimized to balance efficiency and streaming performance. Le et al. \citep{DPCRN} proposed DPCRN, an encoder-decoder with causal two-dimensional convolutions and skip connections. A dual-path module models spectral structure within each frame and temporal dynamics across frames, under causal constraints. Rong et al. proposed GTCRN \citep{GTCRN} as a direct follow-up. It uses grouped temporal convolutions and grouped dual-path recurrent units, merges high-frequency bands through an \textit{equivalent rectangular bandwidth} (ERB) filter bank, and applies temporal recurrent attention. The result keeps the parameter count in the tens of thousands while remaining competitive with heavier systems.
Most recently, Pei et al. proposed aTENNuate \citep{aTENNuate}, a deep state space autoencoder for streaming enhancement on raw audio. The model stacks causal state space blocks with skip connections and light resampling. It targets online operation with steady latency and compact size, and it trains with objectives in both time and frequency domains.
Overall, the field has made a substantial progress from classical-plus-RNN hybrids to highly capable pure \textit{end-to-end} deep neural-network models, for higher intelligibility and naturalness under challenging real-time constraints.

Although a wide array of methods developed in the recent years, including the aforementioned ones, we based our work over \emph{Schröter et al.} \citep{DeepFilterNet2}'s DeepFilterNet2 architecture: a compact two-stage architecture that offers a strong balance of efficiency and quality for streaming SE.
This architecture provides a favorable trade-off between computational complexity and enhancement quality, enabling real-time, low-latency deployment.
%
We chose DeepFilterNet2 as the basis of the proposed model because DPDFNet modifies that backbone directly, preserving its streaming enhancement framework while augmenting the encoder with causal dual-path recurrent blocks. DeepFilterNet3 \cite{DeepFilterNet3} is closely related and may be viewed as a later variant within the same design line, but the architectural intervention proposed here is most naturally described as an extension of DeepFilterNet2.
Motivated by both technical considerations and community interest in advanced speech enhancement, we believe that enhancing the current DeepFilterNet2 architecture will contribute much to the field.

Accordingly, we integrate causal dual-path modules into the DeepFilterNet2 encoder, resulting in \textit{DPDFNet}. The main contribution of this work is a principled extension of the DeepFilterNet2 framework in which causal DPRNN blocks are introduced into its two encoder branches to enrich the convolutional representations with broader temporal and cross-frequency context before fusion, while preserving the original streaming enhancement architecture. In addition, we add an over-attenuation loss, used alongside the original \textit{multi-resolution loss} \citep{DeepFilterNet2}, to mitigate rare cases where the model over-suppresses the target speech, and propose a fine-tuning stage to expose the models to longer-range dependencies. Finally, besides evaluation on the standard VoiceBank+DEMAND and DNS4 blind test benchmarks, we curate a supplementary evaluation set focused on low-SNR conditions with realistic noise types (car, pub, office, etc.) and multilingual coverage spanning 12 languages from the Speech-MASSIVE test set \citep{Speech-MASSIVE}.

The remainder of the paper is organized as follows. Section~\ref{sec:methods} defines the single-channel denoising problem, reviews the DeepFilterNet2 architecture, introduces the causal dual-path RNN block, and describes its integration into the proposed DPDFNet model. Section~\ref{sec:training_framework} details the training framework, including the datasets, augmentation strategy, supplementary evaluation set, and loss functions. Section~\ref{sec:experimental_results} presents the experimental setup and results, including the fine-tuning procedure, the PRISM aggregate metric, OA-loss sensitivity analysis, benchmark comparisons, architectural ablations, and deployment evaluation on Ceva-NeuPro\texttrademark-Nano. Section~\ref{sec:conclusions} concludes the work.

\begin{figure*}
  \centering
  \subfloat[]{%
    \includegraphics[width=0.48\textwidth]{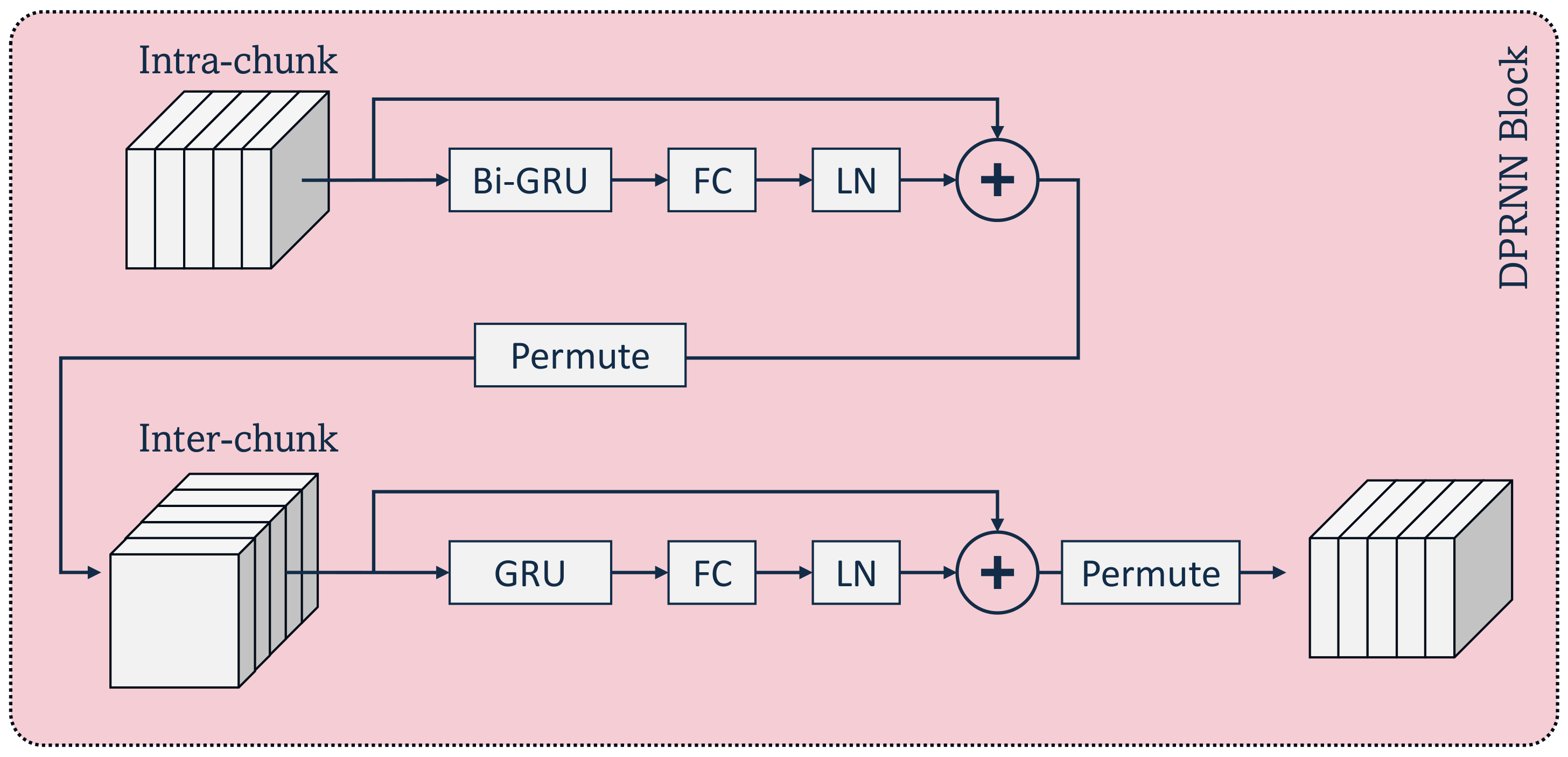}%
    \label{fig:dprnn_block}}
  \hfill
  \subfloat[]{%
    \includegraphics[width=0.48\textwidth]{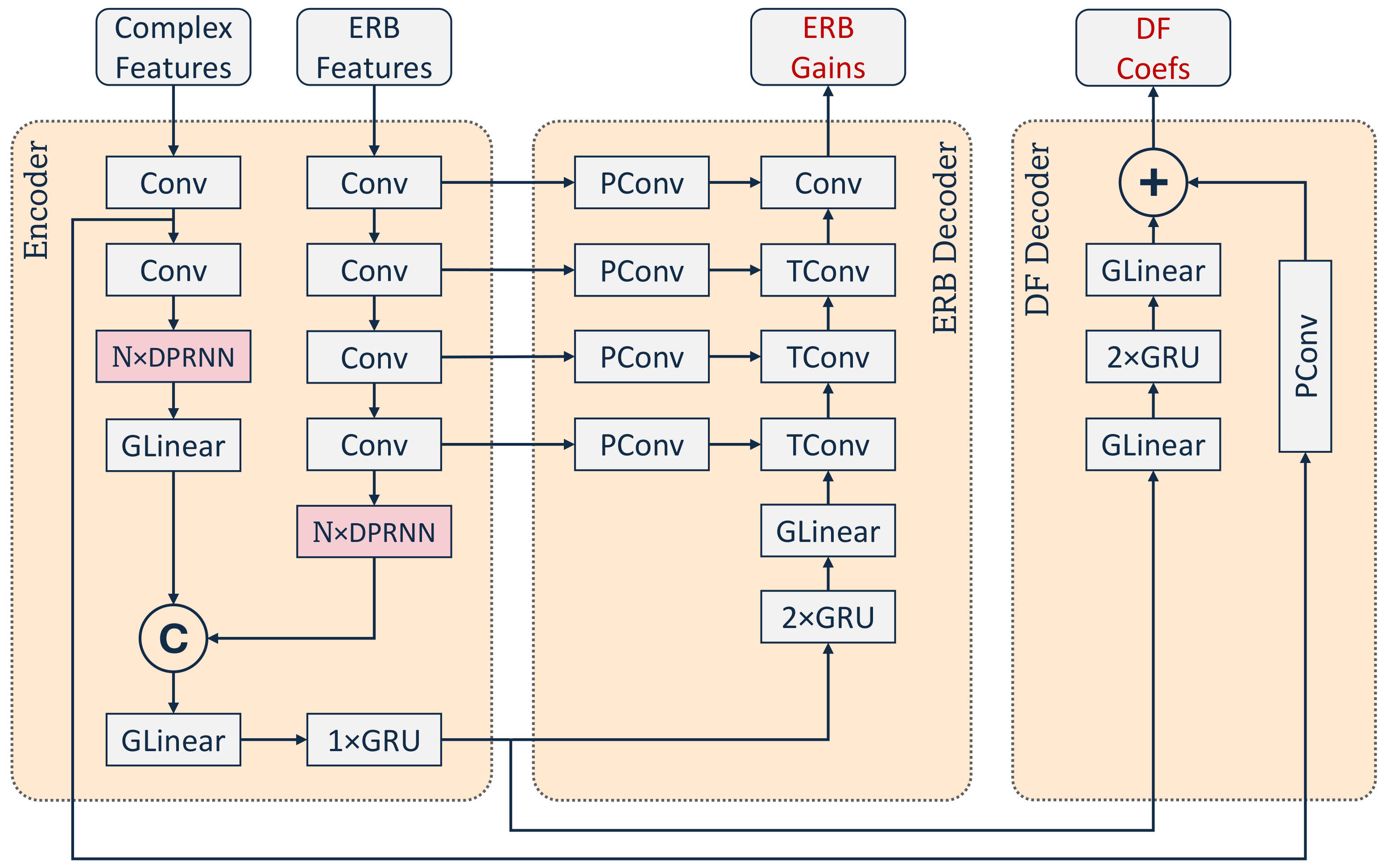}%
    \label{fig:DPDFNet_scheme}}
  \caption{Overview of the proposed DPDFNet architecture. (a) single DPRNN block; (b) DeepFilterNet2 scheme integrated with DPRNN blocks.}
  \label{fig:two_images}
\end{figure*}

\section{Methods}
\label{sec:methods}

\subsection{Denoising Framework}
\label{sec: denoising framework}

We consider a noisy speech signal modeled as
\begin{equation}
    x = s * r + n,
\end{equation}
where $s$ denotes the clean speech signal, $r$ is the room impulse response (RIR) characterizing the acoustic environment, and $n$ represents additive background noise. This formulation captures both the reverberation introduced by the environment and the presence of noise.

Following the original DeepFilterNet2 framework, we apply the Short-Time Fourier Transform (STFT) to the time-domain signal $x$ to obtain its time-frequency representation $X$. This representation is then used to derive two primary input features: \emph{complex features} and \emph{ERB features}.

The complex features consist of the lower $96$ frequency bins of $X$, corresponding to frequencies up to $4800~\text{Hz}$. This range is chosen because it encompasses the majority of the periodic components of speech. We denote these features as $X_{\mathrm{df}}$. In the original DeepFilterNet2, deep-filters (DF) were applied only to these frequency bins.

To obtain the ERB features, the power spectrogram $|X|^2$ is passed through a bank of $32$ ERB filters. This compresses the spectral information in a way that approximates the human auditory system's perception of frequency and energy. The resulting perceptual representation is denoted as $X_{\mathrm{erb}}$.

The denoising framework consists of two sequential stages: \emph{masking} and \emph{reconstruction}.

In the masking stage, the model predicts ERB gains $G_{\mathrm{erb}}(k,b)$, where $k$ denotes the time frame and $b$ denotes the ERB bin index. These gains are applied to the noisy spectrogram as follows:
\\
\begin{equation}
\begin{aligned}
    G(k, f) &= \mathrm{inter}\!\left(G_{\mathrm{erb}}(k, b)\right), \\
    Y_G(k, f) &= X(k, f) \cdot G(k, f),
\end{aligned}
\end{equation}
\\
where $G(k,f)$ is the full-band mask obtained via inverse interpolation of the ERB filter banks (practically, using the transposed ERB filter bank matrix), and $Y_G(k,f)$ is the masked spectrogram.

In the reconstruction stage, the model predicts complex DF coefficients $C(k, i, f_{\mathrm{df}})$, which are applied to the periodic part of the masked spectrogram up to $4800$~Hz:
\begin{equation}
    Y(k, f) = \sum_{i=0}^{N} C(k, i, f) \cdot Y_G(k - i + \ell, f),
\end{equation}
where $N$ is the DF order, $\ell$ is the \emph{look-ahead} ($N=5$ and $\ell=2$ in our case), and the dot product is performed in the complex domain.

\subsection{DeepFilterNet2 Architecture}
\label{sec:deepFilterNet_architecture}
To predict the ERB gains $G_{\mathrm{erb}}(k, b)$ and the DF coefficients $C(k, i, f_{\mathrm{df}})$, the DeepFilterNet2 architecture first uses an \textit{Encoder} that extracts and fuses information from both ERB features and complex features into a unified latent representation, denoted by:
\\
\begin{equation}
\mathcal{E} = \mathcal{F}_{\mathrm{enc}}\left(X_{\mathrm{erb}}, X_{\mathrm{df}}\right)
\end{equation}
This embedding $\mathcal{E}$ then passes through the \textit{ERB Decoder} $\mathcal{F}_{\mathrm{erb\_dec}}$ and the \textit{DF Decoder} $\mathcal{F}_{\mathrm{df\_dec}}$ to predict the ERB gains and DF coefficients, respectively.

More specifically, the \textit{Encoder} consists of two separate branches, each composed of several convolutional blocks. Each convolutional block comprises a separable convolutional layer with 64 channels and a kernel size of $(1, 3)$ (time, frequency), followed by batch normalization and a ReLU activation function. To combine the extracted ERB and complex features, both feature representations are flattened, concatenated, and subsequently processed through \textit{grouped linear} (GLinear) and GRU layers.

The ERB decoder contains two GRU layers, followed by several transposed convolutional blocks, to predict the ERB gains. Each transposed convolutional block comprises a separable transposed convolutional layer, batch normalization, and a ReLU activation, except for the final block, which uses a Sigmoid activation to predict the mask. To establish a U-Net-style architecture, convolutional blocks from the ERB encoder branch are connected to corresponding transposed convolutional blocks in the decoder through skip connections.

The DF decoder consists of two GRU layers, preceded by a GLinear layer and followed by another GLinear layer. Using the output from this structure combined with a skip connection from the encoder, the decoder predicts the complex DF coefficients.

\subsection{Dual-Path RNN Block}
\label{sec:dual_path_rnn_block}
Dual-Path Recurrent Neural Networks, introduced by Luo \etal \citep{DPRNN}, address long-context modeling through two alternating recurrent stages that operate along complementary axes of the data: an \textit{intra stage} that aggregates information within each time frame and an \textit{inter stage} that propagates information across time.

Let the input be a tensor $X \in \mathbb{R}^{B \times T \times F \times D}$, where $B$ denotes batch size, $T$ the number of time frames, $F$ the number of frequency bins, and $D$ the feature dimension.  

\textbf{Intra stage.} To capture spectral dependencies within a time frame, reshape
\[
X \mapsto X_{\text{intra}} \in \mathbb{R}^{B \cdot T \times F \times D}
\]
and apply a bidirectional RNN along the $F$ axis, one sequence per time frame. The recurrent states are initialized to zero independently for each frame-specific sequence. The output is then reshaped back to \(\mathbb{R}^{B \times T \times F \times D}\).

\textbf{Inter stage.} To model temporal evolution at each frequency bin, permute and reshape
\[
X \mapsto X_{\text{inter}} \in \mathbb{R}^{B \cdot F \times T \times D}
\]
and apply a unidirectional RNN along the $T$ axis, one sequence per frequency bin. A single parameter set is shared across all frequencies, while the recurrent states remain distinct. In the causal setting, the hidden states of this inter-stage RNN are initialized to zero. Hence, at the first time frame, the output depends only on the current frame and the zero initial state, while at subsequent frames it depends only on the current and preceding frames, with no access to future context. The result is finally reshaped to \(\mathbb{R}^{B \times T \times F \times D}\).

This dual-path strategy was adopted by DPCRN \citep{DPCRN}, which replaces the GRU bottleneck in a U-Net-style encoder-decoder with DPRNN blocks and reports strong performance. In addition, DPCRN appends a position-wise fully connected (FC) layer and Layer Normalization (LN) \citep{layer_normalization} after each DPRNN stage, both intra and inter, in order to refine features and stabilize training. Figure~\ref{fig:dprnn_block} depicts the scheme of the proposed DPRNN block.

\subsection{DPDFNet}
As outlined in Section~\ref{sec:deepFilterNet_architecture}, DeepFilterNet2 comprises three core components: a shared encoder and two decoders. The encoder takes two complementary feature branches, the ERB features and the Complex features. The ERB branch captures perceptually motivated spectral information, whereas the Complex branch emphasizes periodic structure up to $4800$~Hz, which helps restore the naturalness of human speech.

Before fusion, each branch is processed in parallel by its own stack of convolutional blocks that extract local patterns from the corresponding input. While these convolutions are effective for local time-frequency feature extraction, they remain limited in their ability to model broader temporal dependencies and interactions across frequency bands. The resulting representations are then merged in a combined layer.
To complement this local processing, we insert $N$ DPRNN blocks immediately after the convolutional stacks in each branch. The number of channels in these convolutions is set to the feature dimension $D$. This placement allows each branch to be enriched with more holistic contextual information before fusion, providing richer interactions across time and frequency bins while preserving the original DeepFilterNet2 backbone. The complete DPDFNet architecture is shown in Figure~\ref{fig:DPDFNet_scheme}.

\section{Training Framework}
\label{sec:training_framework}

\subsection{Datasets and Augmentations}
\label{sec:datasets_and_aug}
Many recent SE models, including our baseline DeepFilterNet2, have been trained on the widely adopted Deep Noise Suppression 4 (DNS4) Challenge \citep{dns4} dataset. This full-band (48 kHz) dataset provides 760 hours of clean speech recordings in six languages from 3,230 unique speakers, along with 181 hours of noise comprising 62,000 clips across 150 noise classes. It also includes 248 real and 60,000 synthetic RIRs for generating reverberant signals.
Although DNS4 remains a widely adopted dataset for training SE models, the clean speech portion is not entirely free of residual noise, which may constrain the achievable performance of trained models. In contrast, a variety of high-quality, open-source wideband (16 kHz) datasets exist for both clean speech and noise, offering valuable resources for more effective model training. Furthermore, since the DeepFilterNet2 architecture can be adapted to full-band operation via a lightweight postprocessing step with negligible impact on performance, utilizing these larger and higher-quality wideband datasets presents a particularly compelling direction for training.
Therefore, we downsampled the DNS4 dataset to wideband and enriched it with 4,000 hours of English read speech from the Multilingual LibriSpeech (MLS) corpus \citep{fbmls2020}, a large-scale dataset derived from LibriVox released by Facebook. This addition improves both quality and speaker diversity compared to DNS4 alone. To further ensure the cleanest possible speech signals, all audio was processed through the DPCRN \citep{DPCRN} model. Furthermore, we incorporated noise clips from the MUSAN \citep{musan} and FSD50K \citep{fsd50k} datasets to further diversify the noise conditions.
To improve model generalization, a set of online data augmentations is applied during training. Random second-order filtering is used to expose the model to natural spectral colorations, improving robustness across devices and recording environments. Gain perturbations are introduced to prevent overfitting to specific loudness levels and to ensure invariance to overall volume. In addition, reverberation with early reflection targets is applied, enabling the model to suppress late reverberation while preserving the direct path and early reflections that are beneficial for speech intelligibility.

\subsection{New Evaluation Set}
\label{sec:new_eval_set}
To evaluate and compare different SE models, we report results on the two standard benchmarks most commonly used in single-channel speech enhancement: (1) the VoiceBank+DEMAND test set \citep{vctk_demand} and (2) the DNS4 blind test set \citep{dns4}. In addition, we introduce a complementary multilingual low-SNR evaluation set as a supplementary stress test.

While VoiceBank+DEMAND and DNS4 are well-established and enable direct comparison with prior work, they do not fully cover several conditions relevant to always-on causal SE, such as longer recordings, extended speech-free background-noise segments, lower SNRs, and broader language diversity. Our supplementary set is intended to probe these aspects rather than replace the standard benchmarks.

Accordingly, we developed a supplementary evaluation set designed to better approximate these more challenging real-world scenarios. Specifically, we incorporated nine noisy environments that reflect common day-to-day situations: \textit{airport}, \textit{car}, \textit{office}, \textit{pub}, \textit{rain}, \textit{restaurant}, \textit{street}, \textit{subway}, and \textit{train}. Given the inherently high noise levels in these settings, we selected low SNR values of 0, 5, and 10 dB. To assess cross-linguistic generalization, we used clean speech samples from 12 languages included in the Speech-MASSIVE test set: \textit{Arabic}, \textit{Dutch}, \textit{French}, \textit{German}, \textit{Hungarian}, \textit{Korean}, \textit{Polish}, \textit{Portuguese}, \textit{Russian}, \textit{Spanish}, \textit{Turkish}, and \textit{Vietnamese}. Each clip is approximately 2.5 minutes in duration and features a unique combination of environmental noise, SNR level, and language. Within a single clip, multiple speakers may occur, and speech-free intervals of up to 15 seconds are included to enhance realism. In total, the constructed dataset comprises 324 clips, amounting to roughly 13.5 hours of challenging evaluation material.

\subsection{Loss Functions}
\label{sec:loss_functions}
We adopt the Multi-Resolution loss from  \citep{DeepFilterNet2}. The enhanced signal $y$ is  analyzed with several STFTs, specifically with window sizes $i \in \{5, 10, 20, 40 \}$ ms, to form the following loss: 

\begin{equation}
\mathcal{L}_{\mathrm{MR}}
= \sum_{i} \Big\| \widetilde{Y}_i - \widetilde{S}_i \Big\|_2^{2}
+
\Big\| \widehat{Y}_i - \widehat{S}_i \Big\|_2^{2}
\end{equation}

where the magnitude and phase are represented as 
\begin{align}
\widetilde{Y}_i&=|Y_i|^{c},\quad
\widetilde{S}_i=|S_i|^{c} \\
\widehat{Y}_i&=\widetilde{Y}_i e^{j\phi_{Y_i}},\quad
\widehat{S}_i=\widetilde{S}_i e^{j\phi_{S_i}}
\end{align}

For each resolution $i$, let
\[
Y_i=\mathrm{STFT}_i\{y\},\qquad S_i=\mathrm{STFT}_i\{s\},
\]
denote the complex spectrograms of the enhanced and clean signals, respectively, with phases \(\phi_{Y_i}\) and \(\phi_{S_i}\). The magnitude compression exponent set to $c=0.3$ follow the original DeepFilterNet2 training \citep{DeepFilterNet2}.

During our experiments, we observed that the models suffered to over-attenuation (OA). To address this, we added a loss which penalizes enhanced bins with relatively low energy comparing to the clean target. For each resolution $i$, define a freq-bin binary mask as follows:

\begin{equation}
\mathbf{M}_i(k,f)=\mathbbm{1}\!\left\{ \lvert \mathbf{S}_i\rvert(k,f) > \lvert \mathbf{Y}_i\rvert(k,f) \right\}
\end{equation}

which then applied on the same Multi-Resolution loss:
\begin{equation}
\mathcal{L}_{\mathrm{OA}}
= \sum_{i} \Big\| (\widetilde{Y}_i - \widetilde{S}_i) \odot M_i \Big\|_2^{2}
+
\Big\| (\widehat{Y}_i - \widehat{S}_i) \odot M_i \Big\|_2^{2}
\end{equation}

The total objective combines the both of the lost functions:
\begin{equation}
\mathcal{L} \;=\; \lambda_{\mathrm{MR}}\mathcal{L}_{\mathrm{MR}}+\lambda_{\mathrm{OA}}\mathcal{L}_{\mathrm{OA}} .
\end{equation}
where $\lambda_{\mathrm{MR}} = \lambda_{\mathrm{OA}} = 500$.


A detailed sensitivity analysis of the OA loss, demonstrating its influence on the model’s behavior, is presented in Section~\ref{sec:oa_sensitivity}. Additionally, Table~\ref{tab:overall_results} highlights its importance to the model’s overall performance.

\section{Experimental Results}
\label{sec:experimental_results}

\begin{table*}
    \centering
    \caption{Intrusive, non-intrusive and PRISM scores on our new multilingual low-SNR evaluation set for open-source causal models and our DPDFNet-$\{k\}$ variants ($k$ = number of DPRNN blocks). The baseline uses the DeepFilterNet2 architecture within our training framework.}
    \small
    \setlength{\tabcolsep}{4pt}
    \renewcommand{\arraystretch}{1.15}
    
    \begin{tabular}{lccccccccccccccc}
    \toprule
    \multirow{2}{*}{\textbf{Model}} & 
    \multirow{2}{*}{\textbf{Params}} & 
    \multirow{2}{*}{\textbf{MACs}} & 
    \multirow{2}{*}{\textbf{PESQ}} & 
    \multirow{2}{*}{\textbf{STOI}} & 
    \multirow{2}{*}{\textbf{SI-SNR}} & 
    \multicolumn{4}{c}{\textbf{DNSMOS}} & 
    \multicolumn{5}{c}{\textbf{NISQA}} &
    \multirow{2}{*}{\textbf{PRISM}} \\
    \cmidrule(lr){7-10} \cmidrule(lr){11-15}
     & [M]  & [G]  &  &  &  & \textbf{SIG} & \textbf{BAK} & \textbf{OVL} & \textbf{P.808} & 
     \textbf{MOS} & \textbf{NOI} & \textbf{DIS} & \textbf{COL} & \textbf{LOUD} &  \\
    \midrule
    Noisy & -- & -- & 1.32 & 83.2 & 0.38 & 2.03 & 1.66 & 1.56 & 2.53 & 2.00 & 1.69 & 3.40 & 2.68 & 2.68 & \cellcolor{prismblue}0.04 \\
    \midrule
    DTLN & 0.99 & 0.12 & 2.14 & 88.7 & 10.83 & 2.51 & 3.50 & 2.15 & 2.85 & 2.13 & 2.51 & 2.79 & 2.39 & 2.85 & \cellcolor{prismblue}0.46 \\
    GTCRN & \textbf{0.023} & \textbf{0.039} & 2.21 & 87.2 & 9.11 & 2.56 & 3.76 & 2.25 & 3.00 & 2.53 & 3.27 & 3.17 & 2.80 & 3.01 & \cellcolor{prismblue}0.49 \\
    RNNoise & 0.087 & 0.04 & 2.36 & 88.5 & 8.97 & 2.81 & 3.94 & 2.51 & 3.01 & 1.88 & 3.68 & 2.20 & 2.19 & 3.14 & \cellcolor{prismblue}0.52 \\
    NSNet2 & 2.60 & 0.26 & 2.35 & 86.8 & 8.63 & 2.68 & 3.82 & 2.36 & 2.92 & 2.63 & 3.39 & 3.25 & 2.76 & 3.30 & \cellcolor{prismblue}0.52 \\
    FullSubNet & 5.60 & 30.00 & 2.36 & 89.6 & 10.63 & 2.91 & 3.39 & 2.43 & 3.06 & 2.76 & 2.78 & 3.83 & 3.41 & 3.29 & \cellcolor{prismblue}0.63 \\
    DPCRN & 0.53 & 1.10 & 2.76 & 90.7 & 11.93 & 2.74 & 3.79 & 2.40 & 3.03 & 2.92 & 3.11 & 3.61 & 3.23 & 3.40 & \cellcolor{prismblue}0.69 \\
    aTENNuate & 0.80 & 0.33 & 2.80 & 88.9 & 9.65 & 3.05 & 4.03 & 2.74 & 2.94 & 3.01 & 3.35 & 3.90 & 2.97 & 3.92 & \cellcolor{prismblue}0.70 \\
    DEMUCS & 33.53 & 7.70 & 2.57 & 91.5 & 12.36 & 3.00 & 3.95 & 2.69 & 3.09 & 2.52 & 3.56 & 2.70 & 2.94 & 3.79 & \cellcolor{prismblue}0.72 \\
    DeepFilterNet2 & 2.31 & 0.36 & 2.59 & 91.2 & 11.95 & 2.93 & 3.87 & 2.58 & 3.19 & 3.24 & 3.74 & 3.62 & 3.47 & 3.60 & \cellcolor{prismblue}0.75 \\
    CleanUNet & 46.07 & 15.44 & 2.82 & 93.0 & 12.50 & 3.08 & 4.00 & 2.77 & 3.15 & 2.83 & 3.65 & 2.96 & 3.03 & 3.71 & \cellcolor{prismblue}0.79 \\
    DeepFilterNet3 & 2.14 & 0.35 & 2.76 & 90.5 & 12.10 & 3.03 & 4.01 & 2.72 & 3.22 & 3.84 & 4.12 & 3.99 & 3.75 & 3.85 & \cellcolor{prismblue}0.82 \\
    \midrule
    $\text{Baseline}^*$ & 2.31 & 0.36 & 2.85 & 89.2 & 10.25 & 2.94 & 4.00 & 2.65 & 3.17 & 3.95 & 4.37 & 4.17 & 3.87 & 4.02 & \cellcolor{prismblue}0.79 \\
    \hspace{1em}+OA Loss & 2.31 & 0.36 & 2.92 & 90.2 & 11.21 & \textbf{3.19} & 4.09 & 2.89 & \textbf{3.28} & 3.80 & 4.28 & 4.07 & 3.79 & 4.00 & \cellcolor{prismblue}0.85 \\
    \hspace{2em}+Fine-Tuning & 2.31 & 0.36 & 3.04 & 91.6 & 13.27 & 3.14 & 4.07 & 2.83 & 3.21 & 3.96 & 4.31 & 4.17 & 3.83 & 4.05 & \cellcolor{prismblue}0.91 \\
    DPDFNet-2 & 2.49 & 1.35 & 3.14 & 92.6 & 13.72 & 3.16 & 4.08 & 2.85 & 3.26 & 4.06 & 4.36 & 4.23 & 3.92 & 4.10 & \cellcolor{prismblue}0.95 \\
    DPDFNet-4 & 2.84 & 2.36 & 3.18 & 93.0 & 14.11 & 3.17 & 4.08 & 2.87 & \textbf{3.28} & 4.15 & 4.40 & 4.30 & 3.96 & 4.14 & \cellcolor{prismblue}0.98 \\
    DPDFNet-8 & 3.54 & 4.37 & \textbf{3.20} & \textbf{93.4} & \textbf{14.47} & \textbf{3.19} & \textbf{4.09} & \textbf{2.89} & \textbf{3.28} & \textbf{4.21} & \textbf{4.43} & \textbf{4.34} & \textbf{4.00} & \textbf{4.18} & \cellcolor{prismblue}\textbf{1.00} \\
    \bottomrule
    \multicolumn{16}{l}{$^*$This is the baseline without OA loss and the additional fine-tuning phase.}
    \end{tabular}
    \label{tab:overall_results}
\end{table*}

\begin{table*}
    \centering
    \caption{Objective results on the standard VoiceBank+DEMAND test set for all compared models, including intrusive, non-intrusive, and PRISM metrics.}
    \small
    \setlength{\tabcolsep}{4pt}
    \renewcommand{\arraystretch}{1.15}

    \begin{tabular}{lccccccccccccc}
    \toprule
    \multirow{2}{*}{\textbf{Model}} &
    \multirow{2}{*}{\textbf{PESQ}} &
    \multirow{2}{*}{\textbf{STOI}} &
    \multirow{2}{*}{\textbf{SI-SNR}} &
    \multicolumn{4}{c}{\textbf{DNSMOS}} &
    \multicolumn{5}{c}{\textbf{NISQA}} &
    \multirow{2}{*}{\textbf{PRISM}} \\
    \cmidrule(lr){5-8} \cmidrule(lr){9-13}
     &  &  &  & \textbf{SIG} & \textbf{BAK} & \textbf{OVL} & \textbf{P.808} &
     \textbf{MOS} & \textbf{NOI} & \textbf{DIS} & \textbf{COL} & \textbf{LOUD} & \\
    \midrule
    Noisy & 2.11 & 97.2 & 19.89 & 3.35 & 3.13 & 2.70 & 3.05 & 3.05 & 2.21 & 3.92 & 3.42 & 3.61 & \cellcolor{prismblue}0.23 \\
    \midrule
    NSNet2 & 2.27 & 95.9 & 15.38 & 3.19 & 3.81 & 2.84 & 3.29 & 3.87 & 3.75 & 4.09 & 3.59 & 4.00 & \cellcolor{prismblue}0.25 \\
    RNNoise & 2.22 & 96.7 & 13.84 & 3.29 & 3.86 & 2.95 & 3.39 & 3.45 & 3.85 & 3.63 & 3.36 & 3.94 & \cellcolor{prismblue}0.27 \\
    DTLN & 2.38 & 98.0 & 20.73 & 3.28 & 3.74 & 2.89 & 3.24 & 3.39 & 3.00 & 3.97 & 3.39 & 3.87 & \cellcolor{prismblue}0.44 \\
    DPCRN & 2.37 & 97.8 & 20.88 & 3.25 & 3.71 & 2.84 & 3.36 & 4.05 & 3.67 & 4.25 & 3.88 & 4.17 & \cellcolor{prismblue}0.52 \\
    GTCRN & 2.60 & 97.5 & 21.41 & 3.24 & 3.88 & 2.91 & 3.42 & 4.14 & 3.92 & 4.30 & 3.92 & 4.22 & \cellcolor{prismblue}0.59 \\
    FullSubNet & 2.31 & 98.4 & 17.69 & 3.31 & 3.92 & 3.00 & 3.40 & 4.40 & 4.22 & 4.42 & 4.11 & 4.34 & \cellcolor{prismblue}0.60 \\
    aTENNuate & \textbf{3.06} & 97.9 & 18.33 & 3.41 & 4.06 & 3.14 & 3.38 & 4.00 & 3.91 & 4.18 & 3.63 & 4.11 & \cellcolor{prismblue}0.69 \\
    CleanUNet & 2.59 & 98.7 & 18.16 & 3.42 & 4.07 & 3.16 & 3.48 & 4.18 & 4.11 & 4.21 & 3.84 & 4.21 & \cellcolor{prismblue}0.69 \\
    DEMUCS & 2.89 & \textbf{98.8} & 19.42 & \textbf{3.46} & 4.01 & 3.16 & 3.40 & 3.99 & 3.80 & 4.11 & 3.80 & 4.13 & \cellcolor{prismblue}0.74 \\
    DeepFilterNet3 & 2.80 & 98.2 & 21.22 & 3.44 & 4.10 & 3.19 & 3.58 & 4.60 & 4.36 & 4.60 & 4.18 & 4.44 & \cellcolor{prismblue}0.84 \\
    DeepFilterNet2 & 2.75 & 98.2 & 21.67 & 3.43 & 4.12 & 3.19 & 3.58 & 4.68 & 4.46 & 4.65 & 4.22 & 4.47 & \cellcolor{prismblue}0.85 \\
    \midrule
    Baseline & 2.62 & 98.4 & 22.91 & 3.41 & 4.10 & 3.17 & 3.55 & 4.67 & 4.48 & 4.63 & 4.22 & 4.44 & \cellcolor{prismblue}0.84 \\
    DPDFNet-2 & 2.68 & 98.6 & 23.56 & 3.43 & \textbf{4.14} & 3.21 & 3.60 & 4.73 & 4.51 & 4.68 & 4.26 & 4.48 & \cellcolor{prismblue}0.90 \\
    DPDFNet-4 & 2.66 & \textbf{98.8} & 23.87 & 3.43 & \textbf{4.14} & 3.21 & 3.60 & \textbf{4.78} & \textbf{4.54} & \textbf{4.72} & 4.29 & \textbf{4.51} & \cellcolor{prismblue}0.91 \\
    DPDFNet-8 & 2.71 & \textbf{98.8} & \textbf{24.07} & 3.44 & \textbf{4.14} & \textbf{3.22} & \textbf{3.61} & \textbf{4.78} & \textbf{4.54} & \textbf{4.72} & \textbf{4.30} & \textbf{4.51} & \cellcolor{prismblue}\textbf{0.93} \\
    \bottomrule
    \end{tabular}
    \label{tab:csv_full_results_with_intrusive}
\end{table*}

\begin{table*}
    \centering
    \caption{Objective results on the standard DNS4 blind test set for all compared models, using non-intrusive DNSMOS, NISQA, and PRISM metrics.}
    \small
    \setlength{\tabcolsep}{4pt}
    \renewcommand{\arraystretch}{1.15}

    \begin{tabular}{lcccccccccc}
    \toprule
    \multirow{2}{*}{\textbf{Model}} & 
    \multicolumn{4}{c}{\textbf{DNSMOS}} & 
    \multicolumn{5}{c}{\textbf{NISQA}} &
    \multirow{2}{*}{\textbf{PRISM}} \\
    \cmidrule(lr){2-5} \cmidrule(lr){6-10}
     & \textbf{SIG} & \textbf{BAK} & \textbf{OVL} & \textbf{P.808} & 
     \textbf{MOS} & \textbf{NOI} & \textbf{DIS} & \textbf{COL} & \textbf{LOUD} & \\
    \midrule
    Noisy & 3.23 & 2.41 & 2.26 & 3.03 & 2.09 & 2.12 & 3.40 & 2.71 & 2.72 & \cellcolor{prismblue}0.12 \\
    \midrule
    DTLN & 3.10 & 3.61 & 2.67 & 3.35 & 2.64 & 3.05 & 3.31 & 2.71 & 3.22 & \cellcolor{prismblue}0.40 \\
    RNNoise & 3.20 & 3.94 & 2.89 & 3.52 & 2.46 & 3.66 & 2.89 & 2.57 & 3.21 & \cellcolor{prismblue}0.48 \\
    NSNet2 & 3.08 & 3.78 & 2.72 & 3.46 & 2.88 & 3.42 & 3.48 & 2.89 & 3.36 & \cellcolor{prismblue}0.51 \\
    GTCRN & 3.15 & 3.85 & 2.81 & 3.61 & 3.04 & 3.59 & 3.50 & 3.02 & 3.47 & \cellcolor{prismblue}0.62 \\
    DEMUCS & 3.29 & 4.03 & 3.01 & 3.59 & 2.71 & 3.60 & 3.03 & 2.83 & 3.35 & \cellcolor{prismblue}0.62 \\
    aTENNuate & 3.28 & 3.99 & 2.98 & 3.41 & 2.89 & 3.34 & 3.71 & 2.88 & 3.37 & \cellcolor{prismblue}0.65 \\
    DPCRN & 3.21 & 3.79 & 2.84 & 3.56 & 3.13 & 3.39 & 3.75 & 3.17 & 3.49 & \cellcolor{prismblue}0.67 \\
    CleanUNet & 3.38 & 4.04 & 3.09 & 3.69 & 3.09 & 3.76 & 3.33 & 3.06 & 3.49 & \cellcolor{prismblue}0.78 \\
    FullSubNet & 3.33 & 3.73 & 2.90 & 3.62 & 3.27 & 3.55 & 3.89 & 3.41 & 3.64 & \cellcolor{prismblue}0.80 \\
    DeepFilterNet3 & 3.28 & 3.94 & 2.96 & 3.71 & 3.60 & 4.01 & 3.97 & 3.51 & 3.77 & \cellcolor{prismblue}0.90 \\
    DeepFilterNet2 & 3.32 & 3.99 & 3.02 & 3.73 & \textbf{3.61} & 4.02 & \textbf{3.99} & \textbf{3.54} & 3.75 & \cellcolor{prismblue}0.93 \\
    \midrule
    Baseline & 3.37 & 4.05 & 3.09 & 3.76 & 3.50 & 4.02 & 3.86 & 3.42 & 3.74 & \cellcolor{prismblue}0.94 \\
    DPDFNet-2 & 3.39 & \textbf{4.06} & 3.11 & 3.78 & 3.52 & 4.03 & 3.88 & 3.46 & 3.77 & \cellcolor{prismblue}0.96 \\
    DPDFNet-4 & 3.39 & 4.05 & 3.10 & 3.79 & 3.54 & 4.04 & 3.90 & 3.48 & 3.78 & \cellcolor{prismblue}0.97 \\
    DPDFNet-8 & \textbf{3.40} & \textbf{4.06} & \textbf{3.12} & \textbf{3.80} & 3.53 & \textbf{4.07} & 3.88 & 3.48 & \textbf{3.79} & \cellcolor{prismblue}\textbf{0.98} \\
    \bottomrule
    \end{tabular}
    \label{tab:csv_full_results}
\end{table*}


\subsection{Implementation Details}
\label{sec:impl_details}

We evaluate model performance using a combination of established \textit{intrusive} and \textit{non-intrusive} evaluation metrics. The intrusive metrics include PESQ \citep{rix2001pesq}, STOI \citep{taal2011stoi}, and SI-SNR \citep{leroux2019sdr}, which rely on comparisons between the enhanced signal and a clean, time-aligned reference. For the VoiceBank+DEMAND test set, we observed low-frequency contamination in the provided clean reference signals below 70~Hz. Therefore, when computing the intrusive metrics on this benchmark, we applied a 70~Hz high-pass filter to the signals used for metric evaluation in order to avoid bias from this reference mismatch. The non-intrusive metrics consist of DNSMOS P.835 (SIG, BAK, and OVRL \citep{dnsmosp835}), P.808 MOS \citep{dnsmosp808}, and NISQA~v2.0 \citep{NISQA}, which evaluates Overall Quality, Noisiness, Coloration, Discontinuity, and Loudness.
Audio files were resampled to 16~kHz prior to processing. Training examples were created by segmenting clean speech into 3~sec chunks and mixing them with noise at SNRs randomly sampled from -5 dB to 40 dB. For STFT parameters, we followed the original DeepFilterNet2 setup, employing a window length of 20~ms, a hop size of 10~ms, and a Vorbis window function.
We trained the models for 1.6M iterations with a batch size of 32, employing the AdamW \citep{adamw} optimizer. The learning rate was decayed according to a cosine-annealing schedule, with a pick value of 1e-3.

Among the open-source, we select only those that meet two key criteria: (1) they are causal models suitable for streaming use, and (2) their weights are publicly available, allowing us to run the official versions on our evaluation set. Consequently, all models mentioned in Section~\ref{sec:intro} are included in our evaluations.
For the models introduced in this work, we begin with a \textit{Baseline} model that starts from the same architecture as DeepFilterNet2 \citep{DeepFilterNet2}, retrained within our framework for methodological consistency. We then present three variants of our proposed DPDFNet architecture: \textit{DPDFNet-2}, \textit{DPDFNet-4}, and \textit{DPDFNet-8}. The suffix $k \in {2,4,8}$ indicates the number of DPRNN blocks in the dual-path stack, scaling the model’s capacity from small ($k=2$), to medium ($k=4$), and large ($k=8$).

\begin{figure}
    \centering
    \includegraphics[width=\columnwidth]{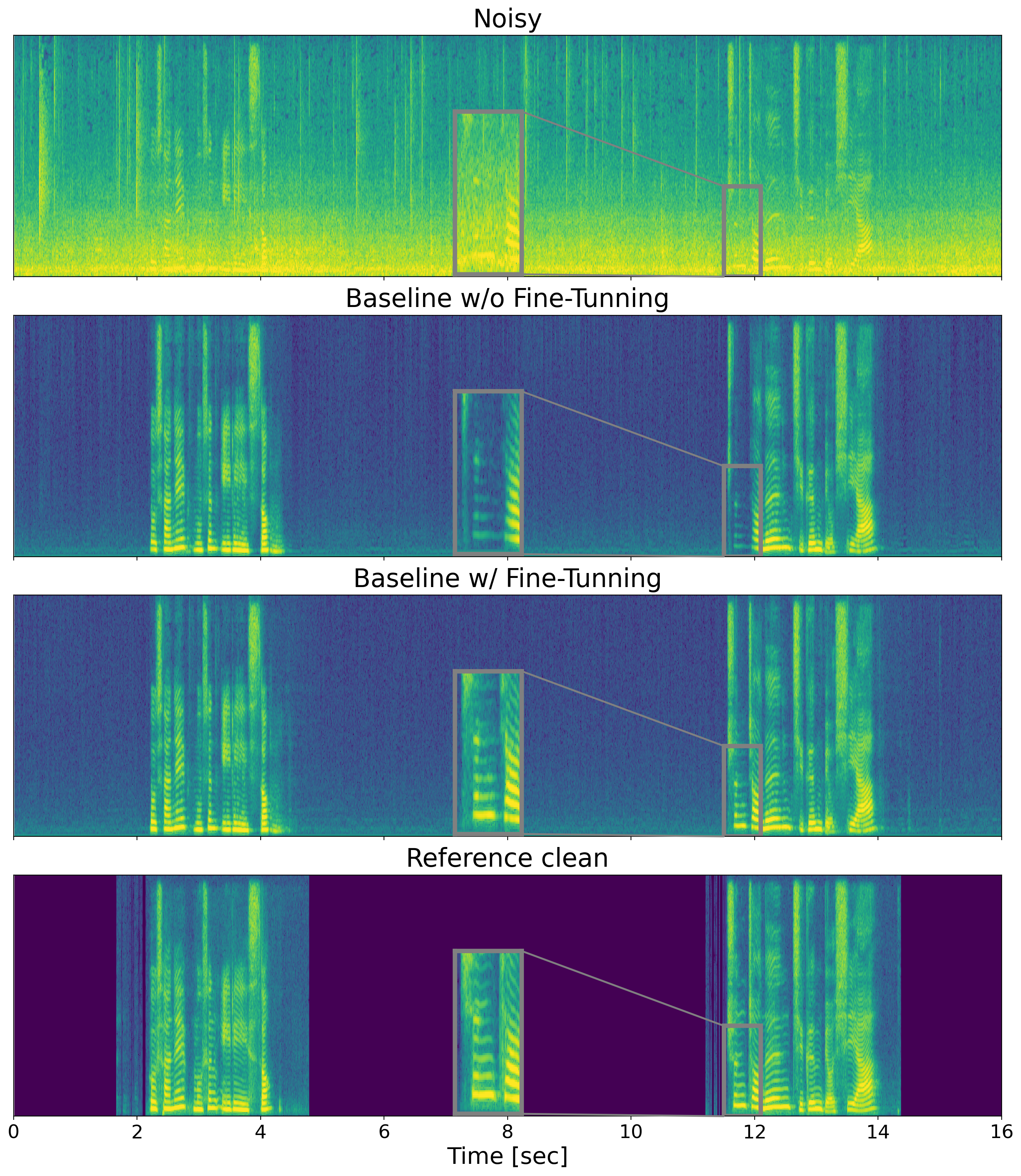}
    \caption{Fine-tuning improves speech recovery after long period of non-speech interval, producing spectrograms closer to the clean reference.}
    \label{fig:fine_tuning_case}
\end{figure}

\subsection{Fine-Tuning}
\label{sec:fine_tuning}
After training the models for 1.6M iterations - covering 42,667 hours of speech - we obtained high-quality speech enhancement systems. However, when deployed in an ``always-on'' streaming setting, we observed instability, including truncation of the initial speech segment following extended silence (3-4~s) and occasional leakage of background noise after such intervals. We hypothesize that these issues arise because stateful components were not sufficiently exposed to state transitions during training, as the models were primarily trained on short, isolated segments.

To address this, we performed an additional fine-tuning stage with batch size of 1, continuous input segments of 30-40~sec, and 5,000 iterations at a fixed learning rate of 1e-5. This procedure encourages the model to better handle long-term temporal dependencies and transitions between speech and non-speech regions encountered in real-time operation.

Following this fine-tuning, all our proposed models exhibited markedly improved stability during continuous inference. A quantitative comparison, with and without this fine-tuning, on our evaluation set is provided in Table~\ref{tab:overall_results}. A representative qualitative example is shown in Fig.~\ref{fig:fine_tuning_case}, where the benefit of the fine-tuning stage is most evident around the transition from extended non-speech segment to speech.

\subsection{Performance Relative Integrated Scaled Metric}
\label{sec:prism}

While the individual metrics mentioned in Section~\ref{sec:impl_details} already able to demonstrate the strength of a SE model, they each emphasize different aspects of enhancement quality. Intrusive measures reward reconstruction fidelity but may undervalue perceptual improvements, whereas non-intrusive predictors better reflect subjective listening impressions but can sometimes favor aggressive noise suppression. As such, viewing each metric in isolation can obscure the holistic trade-off between speech preservation, noise removal, and overall perceptual quality.

To address this limitation, we introduce the \textit{Performance Relative Integrated Scaled Metric} (PRISM).  
PRISM employs a hierarchical scoring framework that integrates both intrusive and non-intrusive objective metrics into a single normalized score. Similar to URGENT~\cite{urgent}, it hierarchically aggregates heterogeneous metrics to reduce ambiguity across multiple evaluation criteria; however, as described below in Section~\ref{sec:prism}, PRISM uses min-max normalization rather than rank-based normalization, thereby preserving the relative magnitude of performance gaps between models.

At the first stage, each metric group Intrusive, DNSMOS P.808 \& P.835, and NISQA is individually normalized using \textit{min-max normalization} across all models, mapping the lowest-performing result to 0 and the highest to 1.  
The normalized metrics are then averaged within each group, forming three composite scores: one for intrusive, one for DNSMOS, and one for NISQA. The DNSMOS and NISQA composites are subsequently combined to represent the non-intrusive category.  

Finally, the overall PRISM score is obtained by taking the mean of the intrusive and non-intrusive composite scores into a unified measure of model performance.  
This hierarchical design provides a more interpretable and balanced aggregation of quality metrics, capturing both signal-based and perceptual performance aspects on a consistent, unified scale.

\begin{figure}
    \centering
    \includegraphics[width=\linewidth]{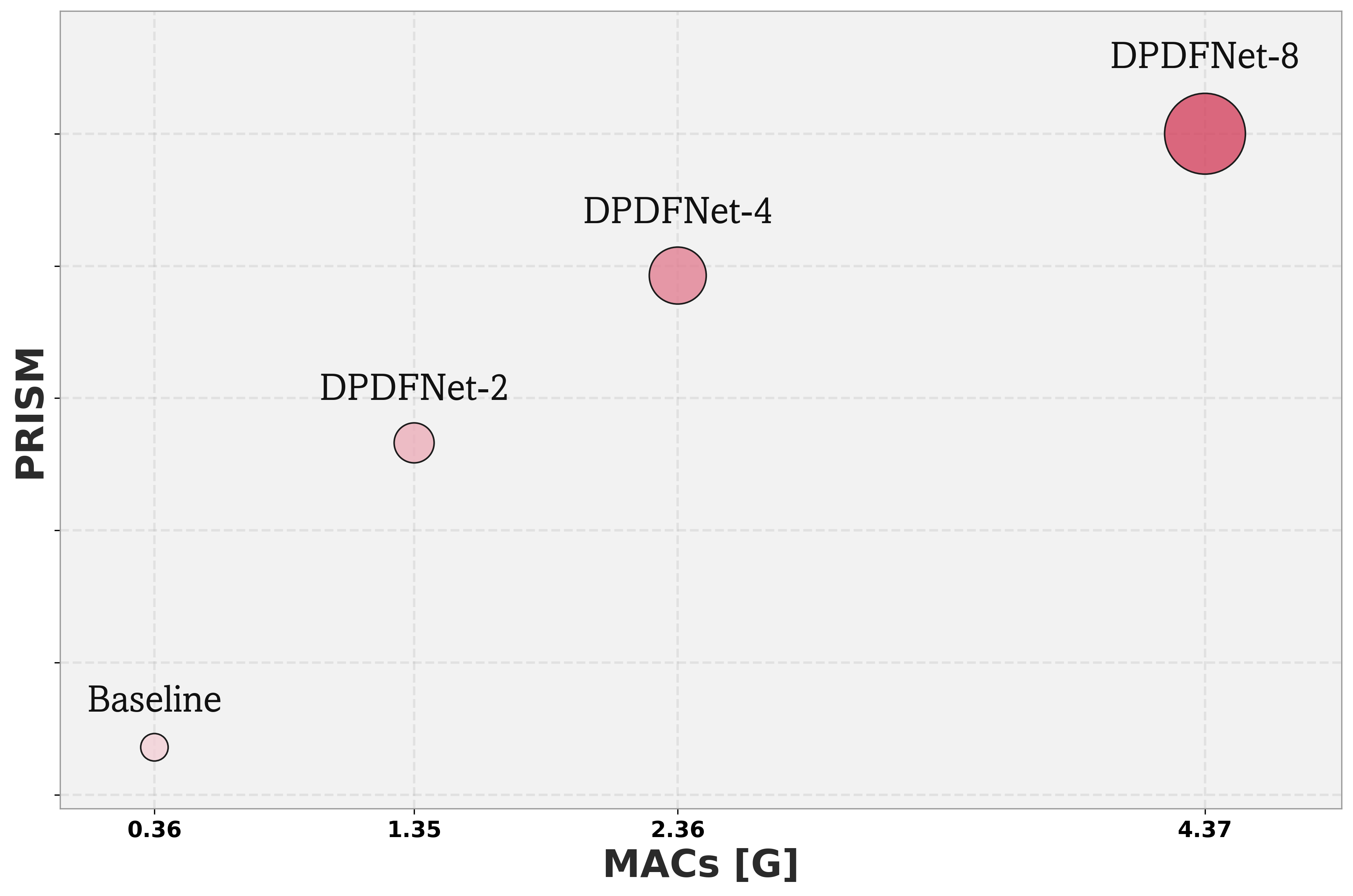}
    \caption{Impact of dual-path block depth in DPDFNet on PRISM performance, showing a direct performance gain with depth, alongside changes in complexity (X-axis) and footprint (bubble size).}
    \label{fig:tradeoff}
\end{figure}

Now that we have a unified metric through PRISM, we can more clearly examine the trade-offs between model quality and efficiency. Figure~\ref{fig:tradeoff} plots PRISM against model complexity (MACs) and bubble size (\#Params). The results reveal a consistent pattern: as the number of DPRNN blocks increases from 0 (\textit{i.e.}, Baseline) to 8, the PRISM score rises monotonically, indicating that quality improvements scale reliably with depth. Meanwhile, computational cost and model size grow only moderately. This offers deployment flexibility: smaller variants are preferable when efficiency is critical, while deeper ones can be chosen when maximizing enhancement quality is the priority.

\subsection{OA-Loss Sensitivity Analysis}
\label{sec:oa_sensitivity}

To better understand the effect of the OA-loss weight, described in Section~\ref{sec:loss_functions}, we perform a dedicated sensitivity analysis using $\lambda_{\mathrm{OA}}\in\{0,500,1000\}$ over the supplementary evaluation set proposed in Section~\ref{sec:new_eval_set}. This analysis is intended to isolate the trade-off controlled by the OA loss:
preserving speech energy while suppressing residual non-speech content. Since standard aggregate enhancement metrics do not explicitly separate these two effects, we define two diagnostic frame-level measures.

Let $r_c(t)$ and $r_e(t)$ denote the short-time RMS values of the clean and enhanced signals at frame $t$, respectively. Specifically, let $\mathcal{S}$ denote the set of frames labeled as speech, and let $\bar{\mathcal{S}}$ denote the set of frames labeled as non-speech.

\paragraph{\textbf{Speech preservation ratio}}
The first metric measures how much clean-speech energy is preserved by the enhancement model in speech-active regions. We define the speech preservation ratio as

\begin{equation}    
\mathrm{SPR}
=
\frac{
\sum_{t \in \mathcal{S}} r_e(t)
}{
\sum_{t \in \mathcal{S}} r_c(t)
}
\end{equation}

Thus, values below 1 indicate attenuation of parts of the speech signal, while values above 1 indicate amplification of parts of the speech signal.

\paragraph{\textbf{Non-speech residual level}}
We define the Non-speech residual level as

\begin{equation} 
\mathrm{NSRL}_{\mathrm{dB}}
=
20 \log_{10}
\left(
\frac{
\frac{1}{|\bar{\mathcal{S}}|}
\sum_{t \in \bar{\mathcal{S}}} r_e(t)
}{
\frac{1}{|\mathcal{S}|}
\sum_{t \in \mathcal{S}} r_c(t)
}
\right)
\end{equation}

Lower values indicate less residual level in non-speech regions relative to the clean speech level, corresponding to stronger suppression of non-speech content.

\begin{figure}
    \centering
    \includegraphics[width=\columnwidth]{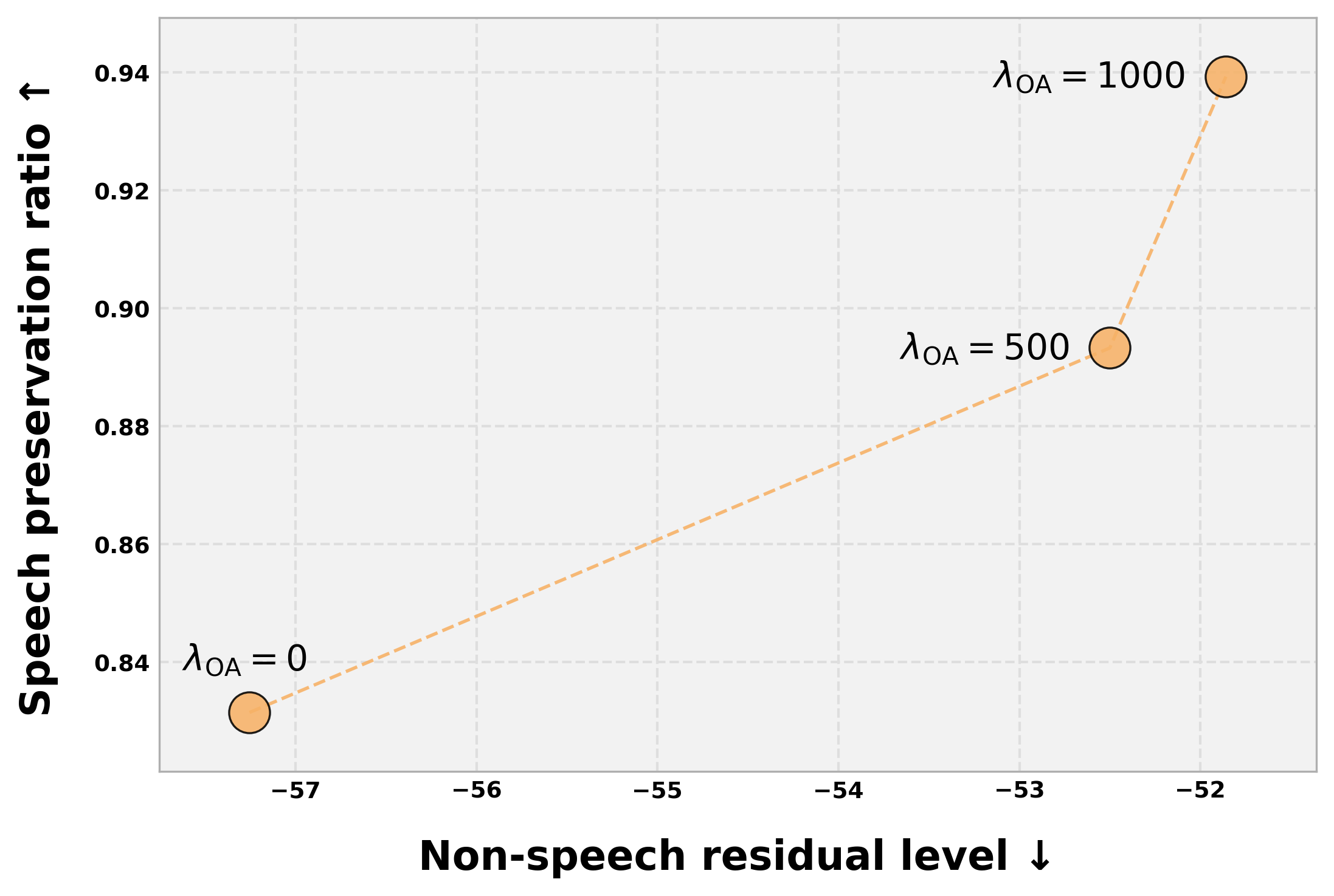}
    \caption{OA-loss sensitivity analysis over the supplementary evaluation set for
    $\lambda_{\mathrm{OA}}\in\{0,500,1000\}$. The vertical axis reports the
    speech preservation ratio, where higher is better. The horizontal axis
    reports the non-speech residual level, where lower is better.}
    \label{fig:oa_sensitivity}
\end{figure}

Figure~\ref{fig:oa_sensitivity} shows the trade-off induced by the OA-loss weight. With $\lambda_{\mathrm{OA}}=0$, the model obtains the lowest non-speech residual level, but also the lowest speech preservation ratio, indicating that speech components are more strongly attenuated together with the non-speech content. Increasing the weight to $\lambda_{\mathrm{OA}}=1000$ improves speech preservation, but also increases the residual level in non-speech regions. The selected value, $\lambda_{\mathrm{OA}}=500$, provides a balanced operating point.

\subsection{Results}
\label{sec:results}

We first turn to our supplementary multilingual low-SNR evaluation set. Table~\ref{tab:overall_results} reports both intrusive and non-intrusive metrics, including reconstruction fidelity measures (PESQ, STOI, SI-SNR) and perceptual quality predictors (DNSMOS P.835 SIG/BAK/OVRL, P.808 MOS, and NISQA). This complementary stress test enables us to assess not only how accurately speech is restored relative to the clean reference, but also how the enhanced signals are expected to be perceived under longer, lower-SNR, and more diverse acoustic conditions.

Several clear trends emerge. The Baseline model - without the OA loss and fine-tuning phase - shows performance comparable to DeepFilterNet2 and DeepFilterNet3. However, once these two training components are added, the resulting Baseline variant surpasses both, indicating that they play an important role in achieving strong enhancement quality.

Turning to our proposed DPDFNet architectures, all variants consistently outperform the Baseline models. Moreover, performance scales positively with the number of blocks (\textit{i.e.}, $k \in \{2,4,8\}$), with DPDFNet-8 achieving the strongest overall results. Compared with strong causal baselines such as DPCRN, and even with substantially larger waveform-domain models such as DEMUCS and CleanUNet, the DPDFNet family remains highly competitive while requiring fewer parameters and lower computational cost.

We then report objective results on the two standard benchmarks used in prior work. Table~\ref{tab:csv_full_results_with_intrusive} presents the VoiceBank+DEMAND results, including intrusive, non-intrusive, and PRISM metrics. Across this benchmark, the DPDFNet variants consistently improve over the Baseline and DeepFilterNet2/3 models, with DPDFNet-8 achieving the strongest overall aggregate performance according to PRISM (0.93). These results show that the proposed dual-path extension improves both signal-level reconstruction and perceptual quality on a widely used standard benchmark.

Table~\ref{tab:csv_full_results} reports the DNS4 blind test-set results. Here as well, DPDFNet improves over DeepFilterNet2/3 and the Baseline in overall perceptual quality, with DPDFNet-8 again achieving the best aggregate PRISM score (0.98). While some competing methods remain slightly stronger on individual submetrics, the overall trend on DNS4 supports the same conclusion as on VoiceBank+DEMAND: adding causal dual-path modeling to DeepFilterNet2 yields a more effective enhancement model overall.

These results show that DPDFNet provides a favorable quality-efficiency tradeoff across both the supplementary evaluation set and standard benchmarks.

\subsection{Architectural Ablation}
To isolate the contribution of the dual-path modules in the two encoder branches, we performed a branch-wise ablation on DPDFNet-2. Starting from the same Baseline architecture, we evaluated two intermediate variants in which the DPRNN stack was inserted only in the Complex branch or only in the ERB branch, while keeping the rest of the model unchanged. The results, reported in Table~\ref{tab:branch_ablation}, show that both single-branch variants improve over the Baseline, confirming that contextual modeling is beneficial in both branches. The gains are larger when the DPRNN is applied only in the Complex branch, indicating that this pathway benefits particularly strongly from the added temporal and cross-band modeling. However, the full DPDFNet-2 model, which applies DPRNN blocks to both branches before fusion, achieves the strongest overall performance, supporting the complementary role of the two branches and the proposed dual-branch design.

\begin{table}
    \centering
    \caption{Branch-wise ablation of DPRNN insertion on the supplementary multilingual low-SNR evaluation set. ``Complex only'' and ``ERB only'' denote variants in which the DPRNN stack is inserted only in the corresponding encoder branch, while ``Both branches'' denotes the full DPDFNet-2 model. Higher is better for all reported metrics.}
    \small
    \setlength{\tabcolsep}{4.5pt}
    \renewcommand{\arraystretch}{1.12}
    \resizebox{\columnwidth}{!}{
    \begin{tabular}{>{\centering\arraybackslash}p{2.4cm}cccc}
        \toprule
        \diagbox[
            width=2.4cm,
            height=2.7em,
            innerleftsep=4pt,
            innerrightsep=4pt,
            linewidth=0.4pt
        ]{\makecell[l]{\textbf{Metric}}}{\makecell[r]{\textbf{Model}}}
        & \textbf{Baseline}
        & \textbf{ERB only}
        & \textbf{Complex only}
        & \textbf{Both branches} \\
        \midrule
        PESQ       & 3.04 & 3.04 & 3.12 & \textbf{3.14} \\
        STOI       & 91.6 & 91.7 & 92.5 & \textbf{92.6} \\
        SI-SNR     & 13.27 & 13.43 & \textbf{13.74} & 13.72 \\
        DNSMOS SIG & 3.14 & 3.13 & 3.14 & \textbf{3.16} \\
        DNSMOS BAK & 4.07 & 4.06 & 4.07 & \textbf{4.08} \\
        DNSMOS OVL & 2.83 & 2.82 & 2.83 & \textbf{2.85} \\
        P.808 MOS  & 3.21 & 3.24 & 3.25 & \textbf{3.26} \\
        NISQA MOS  & 3.96 & 4.02 & 4.05 & \textbf{4.06} \\
        NISQA NOI  & 4.31 & 4.34 & 4.35 & \textbf{4.36} \\
        NISQA DIS  & 4.17 & 4.21 & 4.22 & \textbf{4.23} \\
        NISQA COL  & 3.83 & 3.86 & 3.89 & \textbf{3.92} \\
        NISQA LOUD & 4.05 & 4.09 & \textbf{4.10} & \textbf{4.10} \\
        PRISM      & \cellcolor{prismblue}0.06 & \cellcolor{prismblue}0.26 & \cellcolor{prismblue}0.78 & \cellcolor{prismblue}\textbf{0.99} \\
        \bottomrule
    \end{tabular}}
    \label{tab:branch_ablation}
\end{table}

\begin{figure*}
    \centering
    \includegraphics[width=0.8\linewidth]{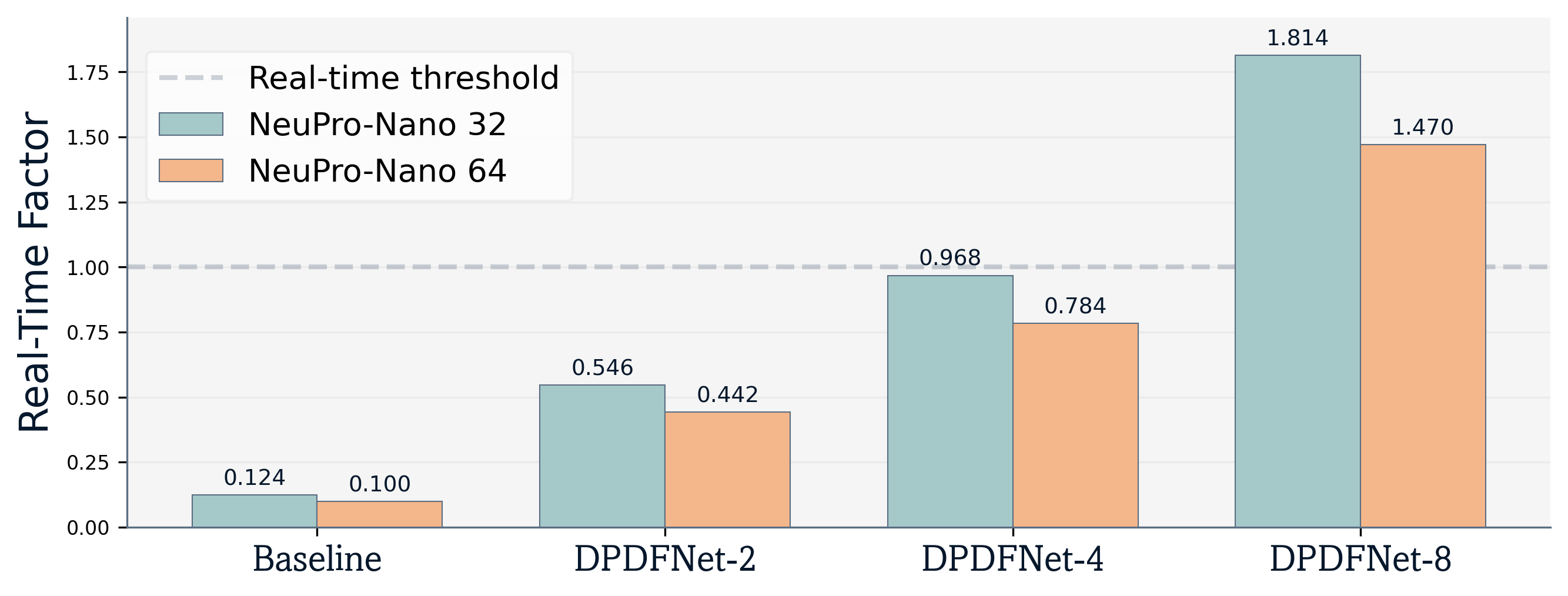}
    \caption{Real-time factor of our proposed models on NPN32 and NPN64 for the target deployment configuration (\textbf{int8} weights and \textbf{int16} activations).}
    \label{fig:real_time_factor}
\end{figure*}

\subsection{Deployment DPDFNet on NeuPro-Nano}

Ceva-NeuPro\texttrademark-Nano\footnote{\url{https://www.ceva-ip.com/product/ceva-neupro-nano/}} (NPN) is a programmable edge NPU IP optimized for embedded deep-learning under stringent power and area constraints. It integrates control, DSP, and neural execution within a single core, removing the need for an external host processor. The architecture supports two configurations (\textit{NPN32} and \textit{NPN64}), integer precisions from 4 to 32 bits, and includes advanced features such as native transformer operators, sparsity acceleration, fast re-quantization, and Ceva’s NetSqueeze\texttrademark ~technology for on-the-fly weight compression to minimize memory traffic.

Meeting real-time (RT) constraints is a key challenge when deploying SE models on hardware with limited memory and computational resources. Many prior works reported RT factors measured on high-performance processors such as the Intel i7. While useful as a baseline, these results do not reflect the conditions of actual edge devices, such as earbuds or smartwatches, where strict power limitations must be considered.
To bridge this gap, we evaluate our models directly on edge-oriented cores, NPN32 and NPN64. The results show that while the largest model, DPDFNet-8, falls short of real-time performance, DPDFNet-4 successfully achieves it on NPN32 with an RT factor of $0.97$. Furthermore, NPN64 delivers a performance improvement, consistently lowering RT factors across all model variants.

These runtime measurements correspond to the target integer deployment configuration used for NeuPro-Nano, consisting of int8 weights and int16 activations. In our deployment flow, this configuration is achieved through quantization-aware training (QAT), rather than by applying naive post-training precision reduction. As the deployment-specific models and toolchain are not included in the public research artifacts, we do not provide a separate public quantization-accuracy study in this paper. Instead, this subsection aims to establish the hardware feasibility and real-time throughput of the proposed approach on edge NPUs under the intended deployment regime.
A detailed comparison of both cores is presented in Figure~\ref{fig:real_time_factor}.

\section{Conclusions}
\label{sec:conclusions}

This work introduced \textit{DPDFNet}, an extension of DeepFilterNet2 that augments the encoder with causal dual-path recurrent blocks to strengthen long-range temporal and cross-band modeling under streaming constraints. We evaluated the proposed models on the standard VoiceBank+DEMAND and DNS4 blind benchmarks, and additionally introduced a supplementary multilingual low-SNR evaluation set covering diverse everyday acoustic scenes. We also proposed PRISM, a scale-normalized composite metric that integrates intrusive and non-intrusive measures.

On the standard VoiceBank+DEMAND and DNS4 blind benchmarks, DPDFNet variants improve over the DeepFilterNet baselines and achieve the strongest overall aggregate performance as measured by PRISM. On our supplementary multilingual low-SNR evaluation set, DPDFNet variants consistently outperform strong causal open-source baselines, including models with substantially higher parameter counts and computational demands. Performance improves monotonically with the depth of the dual-path stack, indicating that the added modeling capacity translates into perceptual gains across both standard and supplementary evaluations.

Finally, we validated embedded feasibility on Ceva-NeuPro\texttrademark-Nano edge NPUs: DPDFNet-4 achieves real-time performance on NPN32 and demonstrates even higher throughput on NPN64.
These results indicate that high-quality, causal single-channel enhancement can be delivered within stringent power and latency budgets typical of on-device applications.

\section*{Data availability}
Data and artifacts supporting the findings of this study are available at:
(1) GitHub repository: \url{https://github.com/ceva-ip/DPDFNet}. 
(2) Hugging Face models: \url{https://huggingface.co/Ceva-IP/DPDFNet}. 
(3) Evaluation set and model outputs: \url{https://huggingface.co/datasets/Ceva-IP/DPDFNet_EvalSet}.
(4) Live demo: \url{https://huggingface.co/spaces/Ceva-IP/DPDFNetDemo}.

\bibliographystyle{elsarticle-harv}
\bibliography{articles}




\end{document}